\documentclass[aip,jcp,reprint,amsmath,amssymb,superscriptaddress,floatfix]{revtex4-2}

\usepackage[T1]{fontenc}
\usepackage[utf8]{inputenc}
\usepackage{graphicx}
\usepackage{bm}
\usepackage{booktabs}
\usepackage{makecell}

\begin{document}

\title{The UZH protocol: Separating errors and constructing improved CP2K basis sets and pseudopotentials}

\author{Hossein Mirhosseini}
\affiliation{Center for Advanced Systems Understanding (CASUS), Conrad-Schiedt-Stra{\ss}e 20, 02826 G\"orlitz, Germany}
\affiliation{Helmholtz-Zentrum Dresden-Rossendorf, Bautzner Landstra{\ss}e 400, 01328 Dresden, Germany}

\author{Tiziano M. A. M\"uller}
\affiliation{Department of Chemistry, University of Zurich, CH-8057 Zurich, Switzerland}

\author{Matthias Krack}
\affiliation{PSI Center for Scientific Computing, Theory and Data, Paul Scherrer Institute, CH-5232 Villigen PSI, Switzerland}

\author{Thomas D. K\"uhne}
\email{tkuehne@cp2k.org}
\affiliation{Center for Advanced Systems Understanding (CASUS), Conrad-Schiedt-Stra{\ss}e 20, 02826 G\"orlitz, Germany}
\affiliation{Helmholtz-Zentrum Dresden-Rossendorf, Bautzner Landstra{\ss}e 400, 01328 Dresden, Germany}
\affiliation{Institute of Artificial Intelligence, Technische Universit\"at Dresden, Helmholtzstra{\ss}e 10, 01069 Dresden, Germany}

\author{J\"urg Hutter}
\affiliation{Department of Chemistry, University of Zurich, CH-8057 Zurich, Switzerland}

\date{\today}

\begin{abstract}
Reliable density-functional simulations require numerical settings whose residual errors are smaller than the chemical and materials trends being interpreted. In CP2K/\textsc{Quickstep}, this requirement is complicated by the joint use of atom-centered Gaussian basis sets and norm-conserving pseudopotentials: a code-to-code discrepancy usually contains both contributions. We present the UZH protocol, a closed-loop CP2K workflow that calibrates molecularly optimized Gaussian basis sets on small molecules, validates the resulting settings in unary-crystal equation-of-state benchmarks, identifies whether the limiting approximation is the Gaussian basis or the pseudopotential. The diagnosis is then used to revise the parameter files. The central diagnostic is a three-way comparison between production CP2K-GTH-UZH calculations, SIRIUS calculations using the same Goedecker--Teter--Hutter pseudopotential in a systematic plane-wave representation, and all-electron full-potential linearized augmented-plane-wave SIRIUS references. This construction decomposes the practical CP2K error into a Gaussian-basis component and a pseudopotential component. The protocol distinguishes basis-limited noble-gas and heavy-element cases from pseudopotential-limited transition-metal cases, guides targeted revisions with the CP2K basis and pseudopotential optimizers, and produces improved MOLOPT basis sets and GTH pseudopotentials as explicit outputs of the workflow. The UZH protocol is therefore constructive: it does not merely measure or reduce errors a posteriori, but allows turning verification outliers into validated CP2K parameter files for simulations across molecules and condensed phases.
\end{abstract}

\maketitle

\section{Introduction}

Density-functional theory is routinely used as a predictive tool in condensed matter physics, chemistry, and materials science~\cite{Kohn1999RMP,Jones2015RMP}. Reproducibility studies have shown that, once exchange-correlation and physical setup are fixed, residual discrepancies between codes are mostly numerical: basis-set completeness, treatment of the core electrons, integration grids, electronic smearing method, and convergence thresholds all matter at the precision level targeted by modern verification workflows~\cite{Lejaeghere2014,Lejaeghere2016,Huber2021,Bosoni2024}. A useful verification protocol therefore has to do more than report agreement with a reference. It should reveal which numerical approximation limits the result and help to convert that information into an improved computational setup.

This issue is particularly important for CP2K/\textsc{Quickstep}~\cite{CP2K2020,CP2KMadeSimple2026}, where Kohn-Sham orbitals are represented by atom-centered Gaussian functions and densities are mapped to auxiliary plane-wave grids within the Gaussian-and-plane-wave (GPW) formalism~\cite{Lippert1997GPW}. The approach is efficient for molecules, liquids, interfaces, and extended systems, but its accuracy is governed jointly by the chosen Gaussian basis sets and norm-conserving pseudopotentials. The two are usually chosen together, and their errors are often reported as a single code-to-code deviation. Such a global error is useful for users, but it is insufficient for method development because it does not say whether the dominant correction should be a larger or reoptimized basis set, a revised pseudopotential, or both. This is precisely where CP2K offers a distinctive advantage as \textsc{Quickstep}, the SIRIUS interface, and the pseudopotential and MOLOPT optimization machinery are available within one coherent code ecosystem, allowing basis-set and pseudopotential errors to be analyzed and improved. Because the production GPW engine, the SIRIUS reference interface, the ATOM pseudopotential generator, and the basis-set optimizer all belong to the CP2K ecosystem, within the UZH protocol one can quantify and minimize Gaussian-basis and pseudopotential errors within the same code framework rather than by stitching together unrelated external tools. Its outcome is consequently not only an error decomposition, but an internally consistent collection of improved MOLOPT basis-set and GTH pseudopotential files.

The aim of the present work is to make this coupled numerical problem actionable: the efficient CP2K/\textsc{Quickstep} production setting is retained, SIRIUS supplies a systematic comparison at fixed pseudopotential and at the all-electron level, and the CP2K basis optimizer and ATOM code provide the route for constructing the improved basis or pseudopotential that the diagnosis identifies as missing.

Here we formulate and apply the UZH protocol as a closed loop between molecular calibration, periodic verification, and parameter improvement. The workflow combines molecularly optimized (MOLOPT) basis sets~\cite{VandeVondele2007} with the analytic Goedecker--Teter--Hutter (GTH) family of separable dual-space Gaussian pseudopotentials and its Hartwigsen--Goedecker--Hutter (HGH) relativistic extension~\cite{Goedecker1996,Hartwigsen1998,Krack2005}. The detailed labels for the three numerical comparisons are introduced in Sec.~II.C. At the conceptual level, the first step asks whether the Gaussian basis is chemically balanced on small molecules, the second step asks how the settings perform in reproducible unary-crystal equation-of-state tests, and the third step identifies which component should be modified. Because CP2K also contains the MOLOPT machinery for basis-set optimization, an atomic code (ATOM) including third-order Douglas--Kroll--Hess (DKH3) calculations~\cite{DouglasKroll1974,Hess1986,NakajimaHirao2000}, and GTH pseudopotential fitting, the same ecosystem can diagnose the limiting approximation and generate revised basis-set or pseudopotential parameter files. The intended product of the protocol is therefore a validated parameter release, not only a retrospective assessment of numerical errors.

\section{Theory and protocol}

\subsection{MOLOPT basis sets}

MOLOPT basis sets are generally contracted Gaussian basis sets designed for density-functional calculations in gas and condensed phases~\cite{VandeVondele2007}. For atom $A$, a contracted basis function in angular-momentum channel $l$ can be written as
\begin{equation}
\chi_{A l m \nu}(\mathbf r)
=
Y_{lm}(\hat{\mathbf r}_A) r_A^{l}
\sum_{p=1}^{N_{Al}} c_{A l\nu p}
\exp\left(-\alpha_{A l p} r_A^2\right),
\label{eq:contracted_basis}
\end{equation}
with $r_A=|\mathbf r-\mathbf R_A|$ and $\hat{\mathbf r}_A=(\mathbf r-\mathbf R_A)/r_A$, where $Y_{lm}$ is a spherical harmonic, $m=-l,\ldots,l$ is the magnetic quantum number, $p=1,\ldots,N_{Al}$ enumerates primitive Gaussians in angular-momentum channel $l$, $\alpha_{A l p}$ are primitive exponents, and $\nu$ labels contractions that differ through the coefficients $c_{A l\nu p}$. Valence and polarization flexibility are therefore introduced mainly through the contraction pattern, while the same primitive pool can be kept sufficiently compact for condensed-phase calculations.

The numerical stability of a candidate basis is measured with the molecular overlap matrix
\begin{equation}
S_{\mu\nu}^{(M)} =\langle \chi_\mu|\chi_\nu\rangle_M
\label{eq:overlap}
\end{equation}
where the composite indices $\mu$ and $\nu$ enumerate contracted basis functions of molecule $M$. For basis level $B$, the condition number is
\begin{equation}
\kappa_B(M)=
\frac{\lambda_{\max}\!\left[S_B^{(M)}\right]}
{\lambda_{\min}\!\left[S_B^{(M)}\right]} ,
\label{eq:condition}
\end{equation}
where $S_B^{(M)}$ is the overlap matrix for basis level $B$, and $\lambda_{\max}$ and $\lambda_{\min}$ denote its largest and smallest eigenvalues. The MOLOPT optimization target balances accuracy and numerical stability, i.e.
\begin{equation}
\Omega =
\sum_{B\in\mathcal B}\sum_{M\in\mathcal T}
\left[
\sum_{q\in\mathcal Q} w_q
\left(
\frac{q_{B,M}-q_M^{\mathrm{ref}}}{s_q}
\right)^2
+ \gamma\ln\kappa_B(M)
\right],
\label{eq:molopt}
\end{equation}
where $\mathcal B$ is the set of related basis levels, $\mathcal T$ is the molecular training set, and $\mathcal Q$ contains the energy and property targets used during calibration. The quantity $q_{B,M}$ is the value obtained with basis level $B$ for molecule $M$, $q_M^{\mathrm{ref}}$ is the corresponding reference value, the scale factors $s_q$ make heterogeneous quantities comparable, the weights $w_q$ control their relative importance, and $\gamma$ weights the condition-number penalty. The logarithmic condition-number term is essential: diffuse primitives improve weak interactions and response properties, but unconstrained diffuse functions can create near-linear dependencies. In the MOLOPT construction, diffuse primitives are included within contractions, providing flexibility without making the overlap matrix numerically fragile.

Two aspects of this construction are important for the UZH protocol. The primitive exponents determine the radial completeness of each angular-momentum channel, whereas the contraction coefficients define compact functions that retain the flexibility needed for valence and polarization response. The condition-number penalty prevents an apparently improved molecular fit from becoming numerically unstable in condensed-phase GPW calculations. Thus, the basis optimization is not merely an enlargement of the basis; it is a constrained optimization of a transferable basis parameter set.

The UZH basis optimization follows this logic but makes the calibration explicit. Small molecules are used first because they probe chemically diverse bonding situations at modest cost, permit comparison to all-electron Gaussian references, and expose failures in geometries, dipoles, polarizabilities, and vibrational frequencies before the settings are transferred to solids. This molecular stage is not the final validation target. It rather establishes a chemically balanced basis set that can then be assessed in a reproducible periodic workflow.

\subsection{Separable dual-space Gaussian pseudopotentials}

The GTH and HGH pseudopotentials are norm-conserving, angular-momentum-dependent pseudopotentials written in an analytic form that is compact in both real and reciprocal space~\cite{Goedecker1996,Hartwigsen1998}. Their local part is
\begin{multline}
V_{\mathrm{loc}}^{\mathrm{pp}}(r) =
-\frac{Z_{\mathrm{ion}}}{r}\operatorname{erf}\left(\alpha^{\mathrm{pp}} r\right)
+ \sum_{i=1}^{4} C_i^{\mathrm{pp}}
\left(\sqrt{2}\alpha^{\mathrm{pp}} r\right)^{2i-2} \\
\times \exp\left[-\left(\alpha^{\mathrm{pp}} r\right)^2\right],
\end{multline}
with $\alpha^{\mathrm{pp}}=1/(\sqrt{2}r_{\mathrm{loc}}^{\mathrm{pp}})$, where $r_{\mathrm{loc}}^{\mathrm{pp}}$ is the local pseudopotential radius. Here $r$ is the electron--nucleus distance, $Z_{\mathrm{ion}}$ is the ionic charge, $i=1,\ldots,4$ enumerates the local Gaussian terms, and $C_i^{\mathrm{pp}}$ are their coefficients. The nonlocal part is written as
\begin{equation}
V_{\mathrm{nl}}^{\mathrm{pp}}(\mathbf r,\mathbf r')
=\sum_{lm}\sum_{ij}
\langle \mathbf r|p_i^{lm}\rangle h_{ij}^{l}
\langle p_j^{lm}|\mathbf r'\rangle,
\end{equation}
where $l$ and $m$ are angular-momentum and magnetic quantum numbers, $i$ and $j$ enumerate nonlocal projectors in a given $l$ channel, $p_i^{lm}$ are projector functions, and $h_{ij}^{l}$ is the channel coupling matrix. The projectors have Gaussian radial factors,
\begin{equation}
\langle \mathbf r|p_i^{lm}\rangle =
N_i^{l}Y^{lm}(\hat r) r^{l+2i-2}
\exp\left[-\frac{1}{2}\left(\frac{r}{r_l}\right)^2\right].
\end{equation}
Here $N_i^l$ is the projector normalization constant, $Y^{lm}$ is the corresponding spherical harmonic, $\hat r=\mathbf r/r$, and $r_l$ is the nonlocal projector radius for channel $l$.
The nonlocal expression above is a Kleinman--Bylander-type separable representation~\cite{Kleinman1982} where each angular-momentum channel is represented by a small set of localized projectors and a low-dimensional coupling matrix $h_{ij}^{l}$ rather than by a fully semilocal radial operator. This form makes the pseudopotential inexpensive to apply in large GPW calculations and directly usable in a plane-wave representation. Norm conservation requires the pseudo-valence states to reproduce the all-electron valence norm and scattering properties outside the core region for each angular-momentum channel. Together with the use of more than one projector per channel, this constraint improves transferability across different bonding environments. The Gaussian radial factors are central to the dual-space construction because they lead to analytic Fourier transforms, so the same GTH parameter set can be evaluated consistently in CP2K/\textsc{Quickstep} and in SIRIUS.

A natural extension within the same GTH family is a nonlinear core correction (NLCC), in which a smooth partial core charge is included in the exchange-correlation functional to account for core-valence density overlap. GTH-NLCC potentials have been constructed by adding a compact Gaussian core charge to the dual-space form and can improve thermochemical transferability, especially for spin-polarized atoms, radical chemistry, and functional transfer~\cite{Willand2013,Li2023GTHNLCC}. We therefore regard NLCC as a compatible extension of the UZH loop: if the SIRIUS-GTH-UZH versus SIRIUS-FP-LAPW comparison identifies a residual pseudopotential error with a core-valence origin, the CP2K ATOM refit can include NLCC parameters in addition to the standard GTH radii, local coefficients, and projector couplings. By contrast, ultrasoft pseudopotential and projector augmented-wave datasets are important alternative frozen-core formalisms in plane-wave electronic-structure codes~\cite{Vanderbilt1990,Blochl1994}, but they are not part of the present decomposition because they would change the pseudopotential representation rather than improve the CP2K GTH/MOLOPT parameter files.

This representation is well suited to GPW calculations because it can be evaluated efficiently with Gaussian orbitals and on real-space grids, while also being compatible with plane-wave calculations. The relativistic HGH generalization extends the construction across the periodic table and retains the separable analytic structure~\cite{Hartwigsen1998}. In the UZH protocol, the GTH form is optimized with CP2K's ATOM program against DKH3 atomic reference calculations, including valence and semicore information as needed.

\subsection{Error decomposition}

We now define the compact method labels used throughout the remainder of the paper. CP2K-GTH-UZH denotes a production CP2K/\textsc{Quickstep} calculation with UZH-MOLOPT Gaussian basis sets and GTH-UZH pseudopotentials. SIRIUS-GTH-UZH denotes the same GTH-UZH pseudopotential evaluated in the SIRIUS plane-wave representation, which removes the CP2K Gaussian-basis approximation. SIRIUS-FP-LAPW denotes the all-electron full-potential linearized augmented-plane-wave reference calculated with SIRIUS. When a targeted parameter revision has been applied, we use the suffix ``improved'' rather than ``new'' to emphasize that the same protocol and data structure are retained.

Let $E_{\mathrm{C}}^{\mathrm{GTH}}(V)$ denote CP2K-GTH-UZH energies, $E_{\mathrm{S}}^{\mathrm{GTH}}(V)$ SIRIUS-GTH-UZH energies, and $E_{\mathrm{S}}^{\mathrm{AE}}(V)$ SIRIUS-FP-LAPW all-electron energies for the same structure and volume, respectively. The total deviation of the practical CP2K protocol is
\begin{equation}
\Delta_{\mathrm{tot}}(V)=
E_{\mathrm{C}}^{\mathrm{GTH}}(V)-E_{\mathrm{S}}^{\mathrm{AE}}(V),
\end{equation}
which can be decomposed as
\begin{equation}
\Delta_{\mathrm{tot}}(V) = \Delta_{\mathrm{basis}}(V)+\Delta_{\mathrm{pp}}(V).\label{eq:decomposition_total}
\end{equation}
The decomposition contains two physically useful terms. The first measures the CP2K Gaussian-basis error for a fixed pseudopotential
\begin{equation}
\Delta_{\mathrm{basis}}(V) =E_{\mathrm{C}}^{\mathrm{GTH}}(V)-E_{\mathrm{S}}^{\mathrm{GTH}}(V), \label{eq:decomposition_basis}
\end{equation}
while the second term measures the pseudopotential error in a systematic basis
\begin{equation}
\Delta_{\mathrm{pp}}(V) =E_{\mathrm{S}}^{\mathrm{GTH}}(V)-E_{\mathrm{S}}^{\mathrm{AE}}(V).\label{eq:decomposition_pp}
\end{equation}

\begin{figure}
\includegraphics[width=\linewidth]{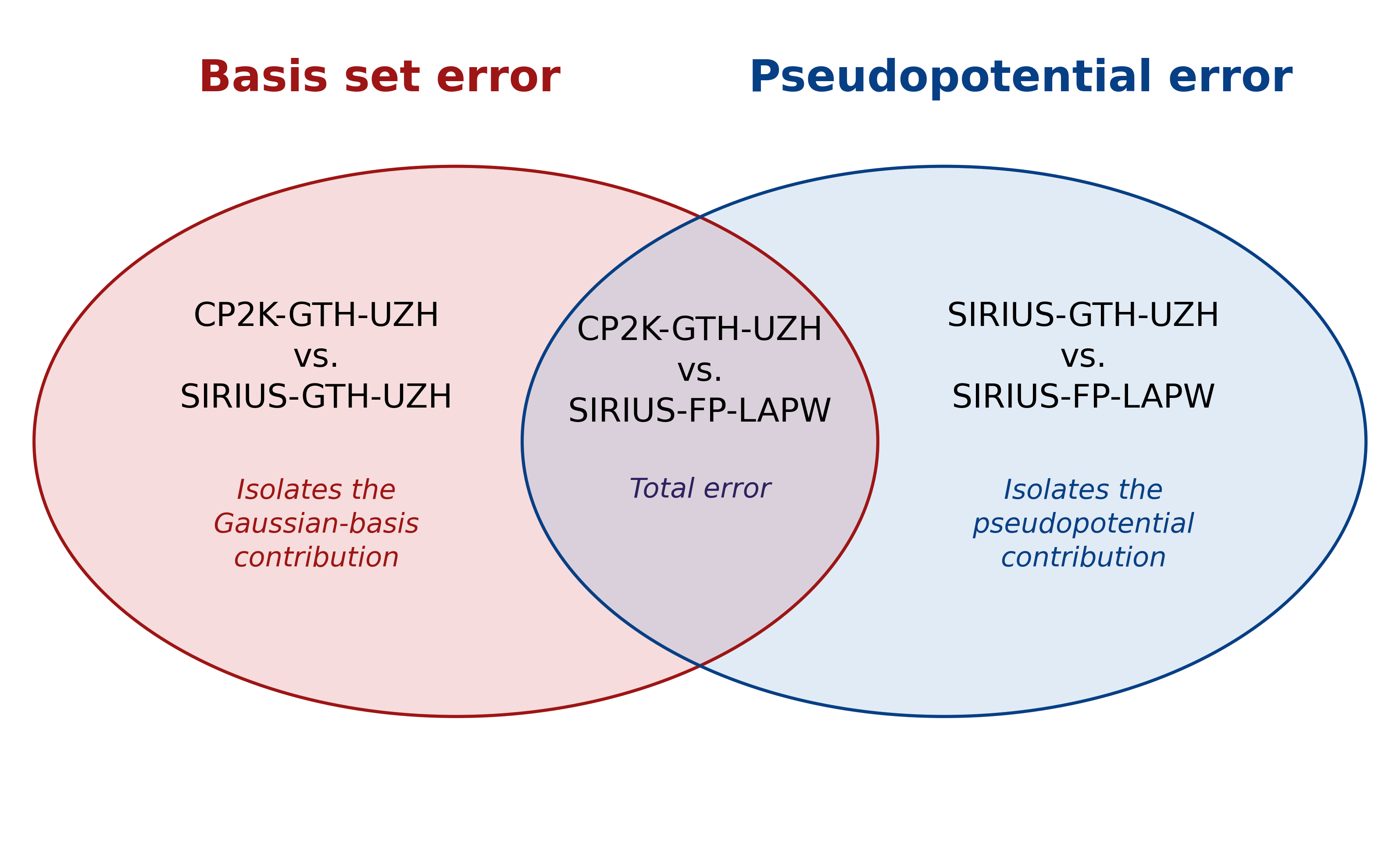}
\caption{\label{fig:decomposition}
Error decomposition in the UZH protocol. CP2K-GTH-UZH compared with SIRIUS-FP-LAPW gives the total practical error. CP2K-GTH-UZH compared with SIRIUS-GTH-UZH isolates the Gaussian-basis contribution. SIRIUS-GTH-UZH compared with SIRIUS-FP-LAPW isolates the pseudopotential contribution.}
\end{figure}

Figure~\ref{fig:decomposition} summarizes this diagnostic construction and shows which pairwise comparison yields the total, Gaussian-basis, and pseudopotential contributions.

In practice, the comparisons are performed not at a single volume but over equation-of-state curves. We use the $\varepsilon$ metric of the verification project~\cite{Bosoni2024},
\begin{equation}
\varepsilon(a,b)=
\sqrt{
\frac{\sum_i \left[E_a(V_i)-E_b(V_i)\right]^2}
{\sqrt{
\sum_i\left[E_a(V_i)-\langle E_a\rangle\right]^2
\sum_i\left[E_b(V_i)-\langle E_b\rangle\right]^2}}},
\label{eq:epsilon}
\end{equation}
where $a$ and $b$ label the two methods being compared, $i$ runs over the volume grid, the volumes $V_i$ are sampled around the equilibrium volume, and $\langle E_a\rangle$ and $\langle E_b\rangle$ are the corresponding mean energies over that grid. This dimensionless metric is sensitive to equation-of-state differences while remaining comparable across chemically distinct elements.

\section{Computational methods}
\subsection{Molecular calibration}
The molecular calibration follows the MOLOPT strategy of optimizing transferable, numerically well-conditioned Gaussian basis sets for chemical accuracy across molecules and condensed phases~\cite{VandeVondele2007}. We benchmarked double-zeta valence plus polarization (DZVP), triple-zeta valence plus polarization (TZVP), and triple-zeta valence plus two polarization shells (TZV2P) UZH-MOLOPT basis sets on a database of small molecules containing diverse bonding motifs; the corresponding CP2K molecular test calculations are archived in the MolTest Zenodo dataset~\cite{HutterMuller2023MolTest}. The molecular tests cover Perdew--Burke--Ernzerhof (PBE), Tao--Perdew--Staroverov--Scuseria (TPSS), and Perdew--Burke--Ernzerhof hybrid (PBE0) exchange-correlation functionals~\cite{Perdew1996,Tao2003TPSS,Adamo1999}. Reference calculations were performed with Gaussian~16~\cite{Gaussian16} using all-electron def2 quadruple-zeta valence plus polarization (def2-QZVP) settings~\cite{Weigend2005}, tight self-consistent-field convergence, and dense numerical integration grids. The CP2K calculations paired all-electron Gaussian-and-augmented-plane-wave (GAPW) calculations~\cite{Lippert1999GAPW} with matching GTH-UZH pseudopotential setups to separate basis and pseudopotential effects at the molecular level. In the molecular figures, DZVP, TZVP, and TZV2P denote CP2K/GPW calculations with GTH-UZH pseudopotentials, whereas the split valence plus polarization (SVP), triple-zeta valence plus two polarization shells (TZVPP), and quadruple-zeta valence plus two polarization shells (QZVPP) entries denote CP2K/GAPW all-electron calculations with corresponding UZH-MOLOPT basis sets, which are also part of the UZH protocol. These all-electron entries are CP2K calculations, not Gaussian~16 calculations; Gaussian~16/def2-QZVP is used as the external molecular reference. For PBE, the available data therefore allow a direct comparison of the GPW/GTH branch with the GAPW all-electron branch, while the TPSS and PBE0 molecular panels shown here emphasize the transferability of the GPW/GTH basis hierarchy. We evaluated relative bond-length errors, maximum angle and torsional errors, dipole moments, average polarizabilities, and vibrational frequencies. For a property $q$, the plotted relative error is $E_R=|q_{\mathrm{CP2K}}-q_{\mathrm{G16}}|/|q_{\mathrm{G16}}|$ unless an absolute angular error is explicitly stated. Structures with incomplete reference or CP2K convergence and structures identified by the frequency analysis as non-minima were excluded from the statistics. Additional details and property-specific statistics are given in the supplementary material~\cite{SuppMat}. Specifically, Sec. I of the supplementary material documents the CP2K pseudopotential optimization using the ATOM code, the CP2K MOLOPT basis-optimization, and SIRIUS setup details needed to reproduce the workflow. The supplementary material Figs.~1--5 collect the molecular calibration statistics that are too extensive for the main text, including angle errors, dihedral errors, dipole moments, bond-class trends, and the Gaussian~16/CP2K all-electron reference comparison.

\subsection{Periodic verification workflow}

Periodic benchmarks were generated with the unary-crystal verification workflow of AiiDA common workflows~\cite{Huber2021,Bosoni2024}. For each element, face-centered cubic (FCC), body-centered cubic (BCC), simple cubic (SC), and diamond prototypes were considered where applicable. Seven volumes were sampled from 94\% to 106\% around the reference volume and fitted to a Birch-Murnaghan equation of state. The same structural inputs and workflow logic were used for CP2K-GTH-UZH, SIRIUS-GTH-UZH, and SIRIUS-FP-LAPW.

CP2K/\textsc{Quickstep} calculations used the GPW method with UZH-MOLOPT basis sets and the corresponding GTH-UZH pseudopotentials. SIRIUS-GTH-UZH calculations used the same GTH-UZH pseudopotentials in a plane-wave representation, thereby removing the Gaussian-basis approximation while retaining the pseudopotential. SIRIUS-FP-LAPW calculations served as the all-electron reference.

SIRIUS is a domain-specific electronic-structure library that implements pseudopotential plane-wave and full-potential linearized augmented-plane-wave methods and is designed for parallel and accelerator-enabled calculations~\cite{Zhang2022Computation,Zhang2023Sirius,SiriusDocs}. In the FP-LAPW mode, the unit cell is partitioned into atom-centered augmentation spheres and an interstitial region; core and valence electrons are treated explicitly, and the density and effective potential are not constrained to a shape approximation. This makes SIRIUS-FP-LAPW an appropriate reference for identifying the residual core approximation in the pseudopotential calculations.

The relativistic treatment was kept consistent with the scalar-relativistic GTH/HGH construction. In the all-electron SIRIUS reference, the full-potential species definitions specify the relativistic treatment of core and valence states; for the FP-LAPW benchmarks reported here, valence scalar relativity is treated with the infinite-order regular approximation (IORA)~\cite{Dyall1999IORA}. This IORA setting is part of the all-electron reference definition, while in the pseudopotential calculations the corresponding DKH3 atomic-reference information is folded into the GTH-UZH potential. Spin-orbit effects are therefore not part of the error decomposition considered here. This choice is deliberate as the objective is to separate Gaussian-basis incompleteness from scalar-relativistic pseudopotential transferability for the CP2K settings used in production GPW calculations. The availability of SIRIUS inside CP2K makes this decomposition especially useful because CP2K can be used both as the production GPW code and as an interface to a systematic plane-wave or all-electron reference. Since SIRIUS is a CP2K component, the verification, optimization, and subsequent improvement loop can be carried out entirely within the CP2K ecosystem rather than by combining unrelated codes.

\subsection{GTH pseudopotential and MOLOPT basis generation}

The first step in generating a consistent set of pseudopotentials and 
basis sets for the periodic table is to choose the separation between 
active (valence and semi-core) and inactive (core) electrons.
For the GTH pseudopotential, a template structure must be selected.
This structure defines the number of nonlocal channels and the form of the local potential. 
The templates may be adapted during the initial optimization cycles, but are typically fixed thereafter.
Pseudopotential optimization is performed against a single all-electron atomic electronic state. 
The reference calculation uses a third-order Douglas--Kroll--Hess scalar relativistic approximation~\cite{CP2K2020,CP2KMadeSimple2026}
The target function is a weighted sum of the differences in occupied and unoccupied orbital energies. 
More details can be found in the supplemental material.

Once the pseudopotentials are available, corresponding MOLOPT basis sets can be generated.
Templates for the basis sets are constructed. The number of primitive functions is chosen 
from test atomic calculations, and the number of contractions for different angular momenta 
follows a double-$\zeta$ polarization, triple-$\zeta$ polarization, triple-$\zeta$ double polarization scheme. 
The same set of small molecules used in the verification step was used for the optimization. 
Reference molecular orbitals for all molecules are generated using a nearly complete basis set.
The MOLOPT basis sets are then optimized against these reference data using a bootstrap procedure. 
First, a combined optimization of hydrogen and first-row elements is performed. 
This is followed by a combined optimization of the second row elements, using the already-optimized sets as needed. 
Finally, all other elements can be optimized based on these two initial sets. 
Optimizations are performed in two steps. First, the exponents are optimized in a double-$\zeta$ polarization setting. 
Then, the contraction coefficients of the complete set are optimized.

More details and additional information can be found in the supplemental material.
In later cycles, if only selected elements are updated, the optimizations can be
restricted. However, if a first or second row element is changed, pseudopotential
or MOLOPT basis, the basis sets for all other elements are affected and have to
be re-optimized.
All optimizations are performed using a Powell-type derivative-free algorithm~\cite{Powell1964}.

\section{Results and discussion}

The results are organized as a diagnostic and constructive sequence that follows the order of the approximations in the protocol. Section~IV.A identifies the smallest converged UZH-MOLOPT basis set suitable for production GPW calculations while retaining numerical stability. Section~IV.B then quantifies the total error of the initial CP2K-GTH-UZH settings before optimization. Section~IV.C separates this total error into Gaussian-basis and pseudopotential contributions by inserting the SIRIUS-GTH-UZH reference between CP2K-GTH-UZH and SIRIUS-FP-LAPW. Finally, Sec.~IV.D evaluates the optimized UZH settings in the same order and shows how the targeted basis-set and pseudopotential revisions become the improved parameter set that reduces the dominant outliers.

\subsection{Molecular basis calibration}

\begin{figure}
\includegraphics[width=\linewidth]{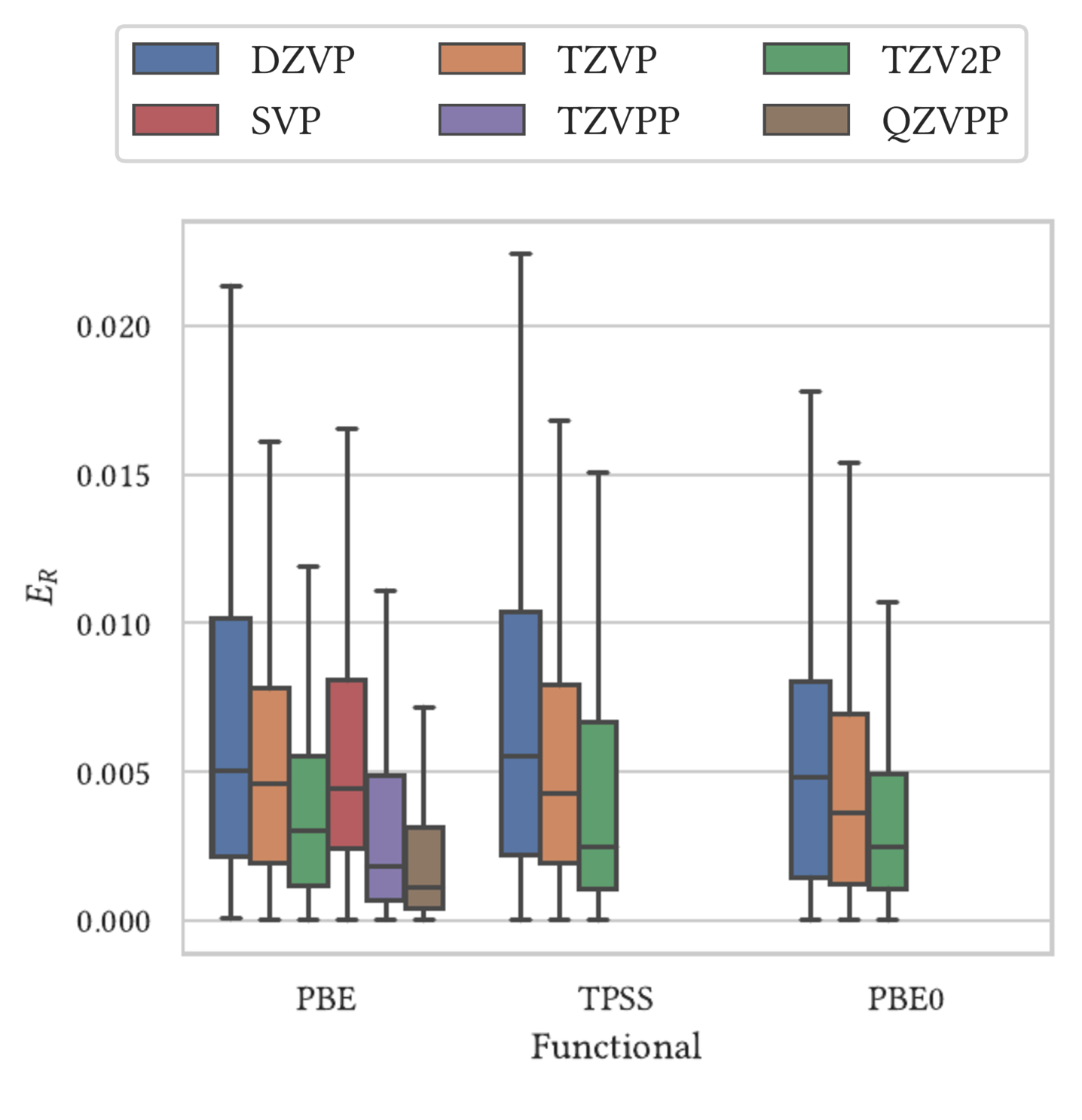}
\caption{\label{fig:bonds}
Relative bond-length errors for the molecular calibration set. The error is $E_R=|d_{\mathrm{CP2K}}-d_{\mathrm{G16}}|/|d_{\mathrm{G16}}|$, where $d_{\mathrm{G16}}$ is the Gaussian~16/def2-QZVP all-electron reference bond length. DZVP, TZVP, and TZV2P are CP2K/GPW calculations with GTH-UZH pseudopotentials. The PBE, TPSS, and PBE0 panels test transferability across exchange-correlation levels. In the PBE panel, SVP, TZVPP, and QZVPP are CP2K/GAPW all-electron calculations with corresponding UZH-MOLOPT basis sets and provide an internal all-electron CP2K comparison. Outliers are omitted to emphasize median and interquartile behavior.}
\end{figure}

The molecular benchmark confirms the expected systematic behavior of the UZH-MOLOPT sequence. Figure~\ref{fig:bonds} reports the first and most local test: relative bond-length errors with respect to the all-electron molecular reference. The boxes measure the relative distance from the Gaussian~16/def2-QZVP reference, $E_R=|d_{\mathrm{CP2K}}-d_{\mathrm{G16}}|/|d_{\mathrm{G16}}|$, where $d$ is a matched bond length. The distribution narrows from DZVP to TZVP and TZV2P, with the median relative error remaining small across the database. This trend is visible for PBE, TPSS, and PBE0, showing that the GPW/GTH-UZH hierarchy is not tuned to a single exchange-correlation functional. The PBE panel additionally contains the UZH all-electron GAPW basis families SVP, TZVPP, and QZVPP, which provide an internal CP2K all-electron comparison for the same molecular set. Carbon-containing bonds are particularly stable, while O- and F-containing compounds are more sensitive to polarization and diffuse flexibility in the smaller basis sets. The TZV2P level removes most of these systematic trends.

This molecular step is not intended to replace the periodic verification workflow, but it is the chemically local calibration that makes the subsequent solid-state test meaningful. A basis that performs well for equation-of-state curves but gives unbalanced molecular geometries, dipoles, or polarizabilities would not be transferable. Conversely, a molecularly optimized basis still has to be checked in condensed phases because diffuse functions, near-linear dependencies, and grid-coupling errors can become more severe in periodic calculations. The UZH protocol therefore uses the molecular benchmark as a first filter: the basis sequence must improve monotonically for local structural observables while keeping the overlap matrix sufficiently well conditioned for production GPW calculations.

Bond angles and dihedral angles extend the same analysis beyond local bond distances. They are more sensitive than bond lengths to subtle changes in the electronic structure and to the automated matching of nearly symmetric structures, but the hierarchy of basis quality remains visible; supplementary material Figs.~1 and 2 report the full angle and dihedral-error distributions, respectively.

The response-property tests complete the molecular calibration. Dipole moments follow the same systematic basis set trend with the full distribution given in supplementary material Fig.~3. Figure~\ref{fig:molecular_polarizabilities} shows the average molecular polarizability, the most demanding response observable in this set. The larger variance observed for polarizabilities is consistent with the need for diffuse flexibility, but the absence of catastrophic outliers at the TZV2P level indicates that the condition-number constraint in Eq.~(\ref{eq:molopt}) is effective. The supplemental bond-class and reference-code comparisons in supplementary material Figs.~4 and 5 further support this choice.
They show that the TZV2P improvement is not confined to the aggregate bond-length distribution and that the molecular calibration is not dominated by a single reference-code mismatch.

\begin{figure}
\centering
\includegraphics[width=\linewidth]{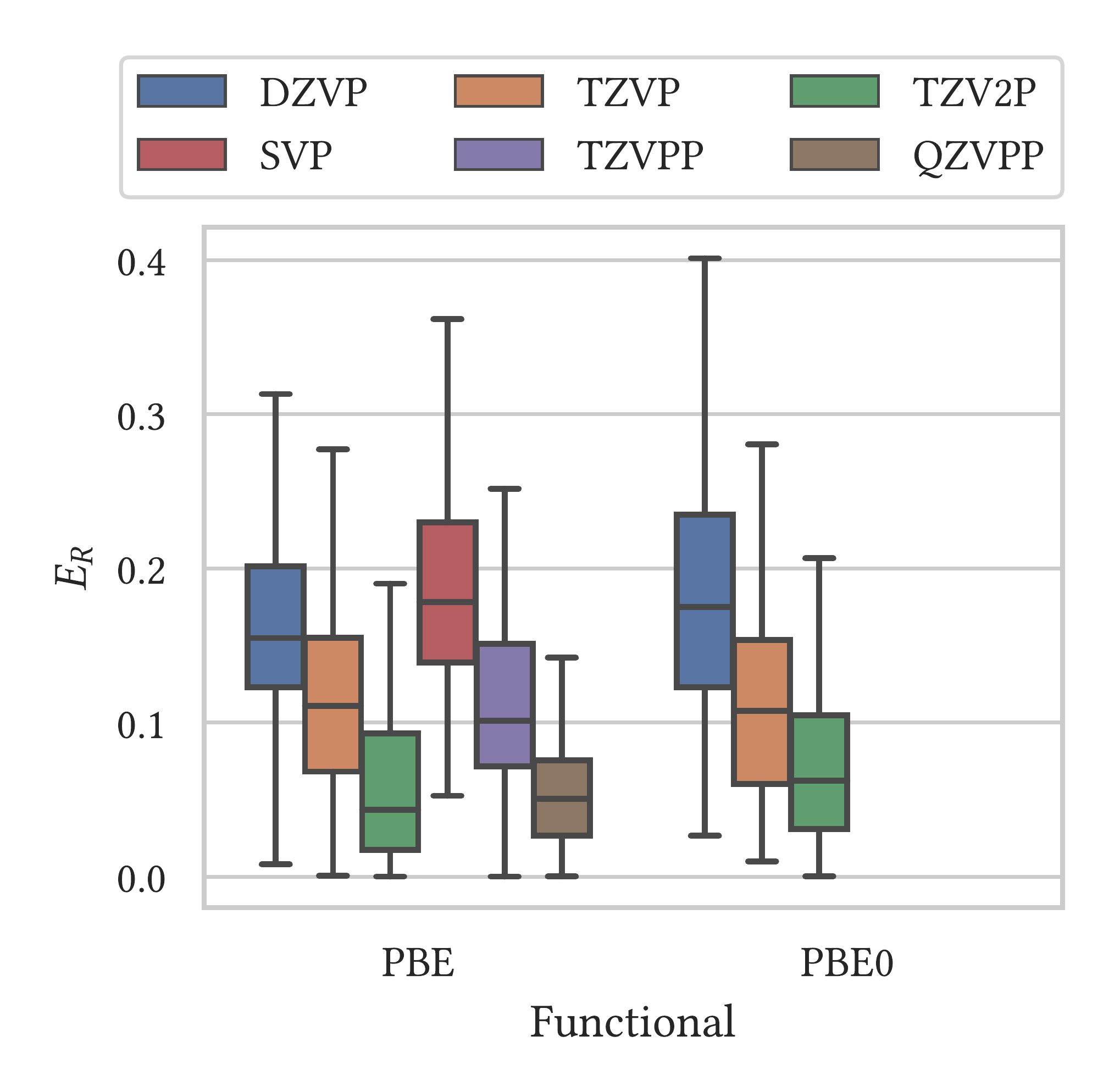}
\caption{\label{fig:molecular_polarizabilities}
Relative errors for average molecular polarizabilities, $E_R=|\bar{\alpha}_{\mathrm{CP2K}}-\bar{\alpha}_{\mathrm{G16}}|/|\bar{\alpha}_{\mathrm{G16}}|$, for the PBE and PBE0 subsets. Polarizabilities are the most demanding molecular response test in this set because they require sufficient diffuse flexibility; TPSS data are omitted because this property was not available for the corresponding CP2K calculations.}
\end{figure}

Together, the bond-length and polarizability benchmarks, complemented by the angular, dihedral, and dipole-moment statistics in supplementary material Figs.~1--3, motivate two practical choices. First, DZVP is useful as an economical screening level, but it should not be interpreted as a converged basis for response-sensitive properties. Second, TZV2P provides a more balanced reference level for the UZH optimization loop because it improves both covalent and polar bonding motifs without introducing the numerical fragility that would compromise large periodic calculations. This balance between chemical flexibility and numerical stability is essential. The goal is not to make the largest possible Gaussian basis, but to construct a basis that can be improved systematically together with the corresponding GTH-UZH pseudopotential files. Consequently, all periodic CP2K-GTH-UZH calculations discussed below use the TZV2P UZH-MOLOPT basis level unless stated otherwise. A complementary all-electron branch of the UZH protocol, based on GAPW MOLOPT basis families and using the same molecular and equation-of-state benchmarks, will be reported separately.

\subsection{Total error before and after improvement}

\begin{figure*}
\includegraphics[width=\textwidth,height=0.88\textheight,keepaspectratio]{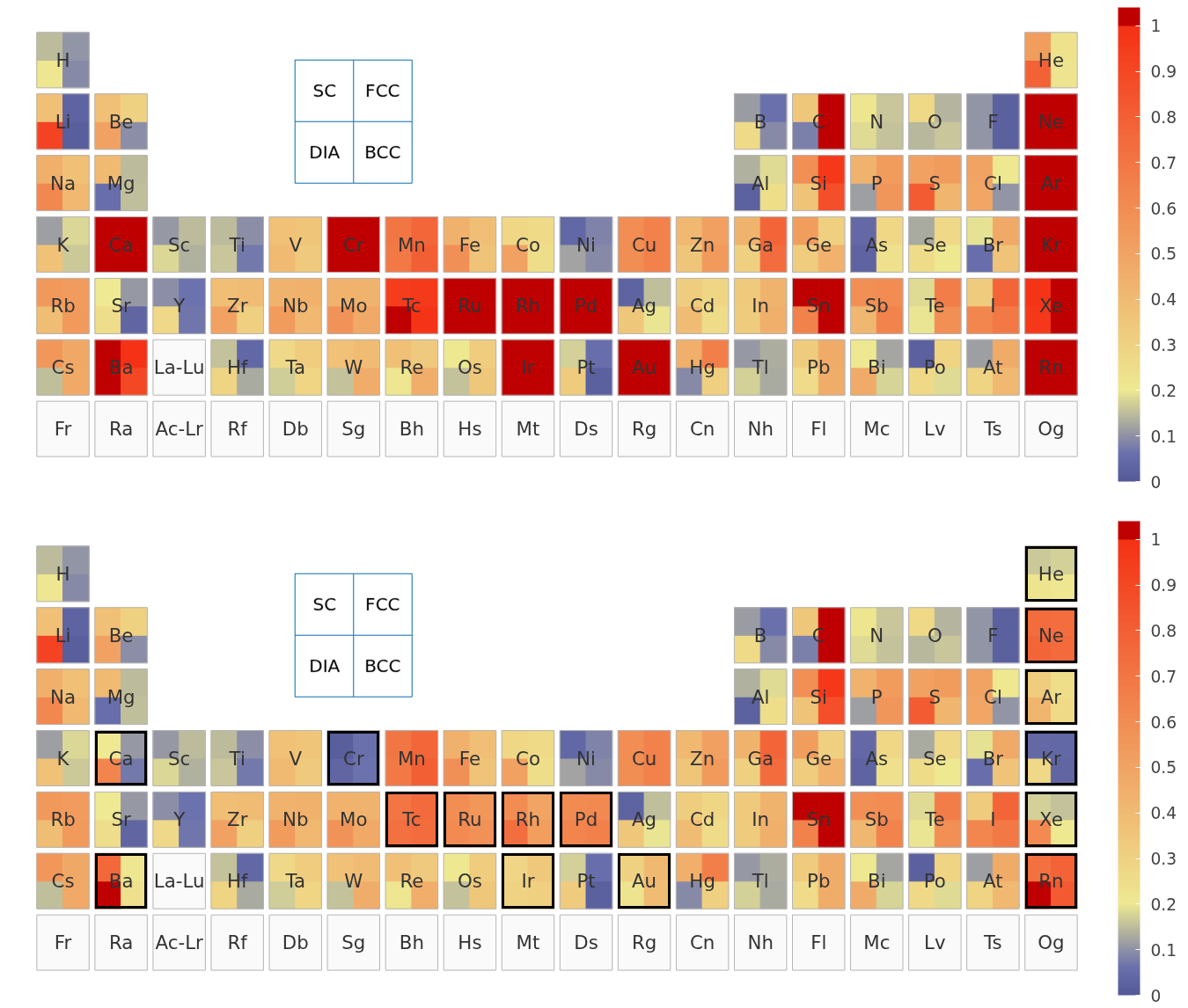}
\caption{\label{fig:total_before_after}
Total equation-of-state error relative to SIRIUS-FP-LAPW, expressed with the $\varepsilon$ metric of Eq.~(\ref{eq:epsilon}). Top: initial CP2K-GTH-UZH settings. Bottom: completed view in which entries absent from the improved data set are filled with the corresponding initial CP2K-GTH-UZH values; black frames mark the recomputed improved entries. Each map contains both Gaussian-basis and pseudopotential errors and therefore represents the practical production-level CP2K error before and after applying the UZH protocol.}
\end{figure*}

Figure~\ref{fig:total_before_after} compares the total practical error of the initial and improved CP2K-GTH-UZH settings relative to the SIRIUS-FP-LAPW reference. In the initial settings, several elements exhibit $\varepsilon>1$ for at least two unary structures, including noble gases, some transition metals, and heavier main-group elements. After applying the UZH protocol, the corresponding CP2K-GTH-UZH-improved map is substantially cleaner, with the largest residual deviations confined to a smaller set of chemically interpretable cases. The completed map in the lower panel fills entries not recomputed with the improved settings by retaining the corresponding initial values, providing a full-period overview on the same scale; the recomputed improved entries are highlighted by black frames. Table~\ref{tab:epsilon-cp2k-uzh-merged-vs-sirius-fp-lapw} gives the corresponding structure-resolved numbers for the recomputed elements. It shows that the improvement is not restricted to a single prototype: noble gases, Cr, Ru, Rh, Pd, Ir, and Au move from values around or above unity to substantially smaller errors for most FCC, BCC, SC, and diamond tests, while the remaining large Ba and Rn diamond entries identify cases that still require additional scrutiny. This before--after comparison is the error a user would observe if CP2K were treated as a black box. It is therefore the most direct measure of the protocol's practical accuracy, but by itself it does not reveal the origin of the discrepancy.

The important point is that the outliers do not have a single chemical or numerical origin. Noble-gas solids are especially sensitive to the radial flexibility and diffuseness of the valence basis because their equation-of-state curves are shallow and dominated by weak interactions. Several transition-metal cases, in contrast, point to the quality of the frozen-core representation, semicore treatment, and the scalar-relativistic transferability encoded in the DKH3-fitted pseudopotential. The total-error map is therefore deliberately used as a starting point rather than as a final ranking of elements: it identifies where the initial settings are insufficient, but the remedial action requires the decomposition in the next subsection.

This distinction is central for interpreting $\varepsilon$. A large total error does not automatically mean that the CP2K Gaussian basis must be enlarged, and a small total error does not prove that basis and pseudopotential errors are individually negligible as partial error cancelation is possible. For parameter development, the total CP2K-GTH-UZH versus SIRIUS-FP-LAPW comparison in Fig.~\ref{fig:total_before_after} has to be read together with the two intermediate comparisons distilled in Table~\ref{tab:classification}. Table~\ref{tab:classification} should therefore be read as the decision table for the first optimization pass: it converts the two pairwise error maps into the component that should be changed. Only then can the protocol decide whether the next iteration should modify the MOLOPT basis, refit the GTH pseudopotential parameters, or treat both components.

\begin{table*}[htbp]
\centering
\caption{Structure-resolved $\varepsilon$ values for CP2K-GTH-UZH relative to SIRIUS-FP-LAPW for the elements recomputed after improvement. Each prototype reports the initial and improved settings side by side; values below unity indicate substantially improved equation-of-state agreement on the verification scale used here.}
\label{tab:epsilon-cp2k-uzh-merged-vs-sirius-fp-lapw}
\begin{tabular}{lrr@{\hspace{1.55em}}rr@{\hspace{1.55em}}rr@{\hspace{1.55em}}rr}
\hline
Element & \multicolumn{2}{c}{Diamond} & \multicolumn{2}{c}{FCC} & \multicolumn{2}{c}{BCC} & \multicolumn{2}{c}{SC} \\
 & initial & improved & initial & improved & initial & improved & initial & improved \\
\hline
He & 0.7904 & 0.2258 & 0.2356 & 0.1744 & 0.2216 & 0.2122 & 0.5329 & 0.1665 \\
Ne & 1.4836 & 0.7737 & 1.7451 & 0.7406 & 1.7018 & 0.7519 & 1.5810 & 0.7450 \\
Ar & 1.3158 & 0.4301 & 1.6646 & 0.2498 & 1.5667 & 0.2650 & 1.4525 & 0.3211 \\
Ca & 1.0954 & 0.6364 & 1.1562 & 0.1080 & 1.3434 & 0.0705 & 1.7059 & 0.2018 \\
Cr & 1.0475 & 0.0333 & 1.2951 & 0.0598 & 1.3602 & 0.0628 & 1.2515 & 0.0157 \\
Kr & 1.7388 & 0.2758 & 1.8142 & 0.0375 & 1.7521 & 0.0337 & 1.8303 & 0.0429 \\
Tc & 1.0057 & 0.7302 & 0.9678 & 0.7530 & 0.9937 & 0.7458 & 0.9554 & 0.7055 \\
Ru & 1.6774 & 0.6156 & 1.5582 & 0.5594 & 1.5601 & 0.5823 & 1.6216 & 0.6050 \\
Rh & 1.6477 & 0.7382 & 1.5784 & 0.5017 & 1.5789 & 0.5258 & 1.5760 & 0.6071 \\
Pd & 1.6345 & 0.6413 & 1.4643 & 0.6186 & 1.4594 & 0.6604 & 1.5064 & 0.6105 \\
Xe & 0.9720 & 0.6154 & 1.3670 & 0.1580 & 1.3011 & 0.2052 & 0.9874 & 0.1723 \\
Ba & 1.8967 & 1.5912 & 0.9984 & 0.2134 & 0.9087 & 0.2399 & 1.4169 & 0.7621 \\
Ir & 1.5403 & 0.2984 & 1.2250 & 0.3439 & 1.1608 & 0.3148 & 1.2696 & 0.2879 \\
Au & 1.6294 & 0.2256 & 1.4091 & 0.4089 & 1.4188 & 0.3981 & 1.4754 & 0.3052 \\
Rn & 1.5540 & 1.2897 & 1.6896 & 0.7849 & 1.6271 & 0.8211 & 1.7891 & 0.7240 \\
\hline
\end{tabular}
\end{table*}

\begin{table*}
\caption{\label{tab:classification}
Classification of dominant error sources in the initial UZH settings. The assignment is based on the pairwise comparisons in Figs.~\ref{fig:total_before_after} and \ref{fig:pp_basis_old} and identifies whether the first optimization pass should target the Gaussian basis or the GTH pseudopotential.}
\begin{ruledtabular}
\begin{tabular}{lll}
Dominant source & Diagnostic signature & Representative elements \\
\hline
Basis set & Large CP2K-GTH-UZH vs SIRIUS-GTH-UZH error & He, Ne, Ar, Kr, Xe, Rn, Ba, Au, Ir \\
Pseudopotential & Large SIRIUS-GTH-UZH vs SIRIUS-FP-LAPW error & Ca, Cr, Pd, Rh, Ru, Tc \\
\end{tabular}
\end{ruledtabular}
\end{table*}

\subsection{Separating basis and pseudopotential errors}

\begin{figure*}
\includegraphics[width=\textwidth]{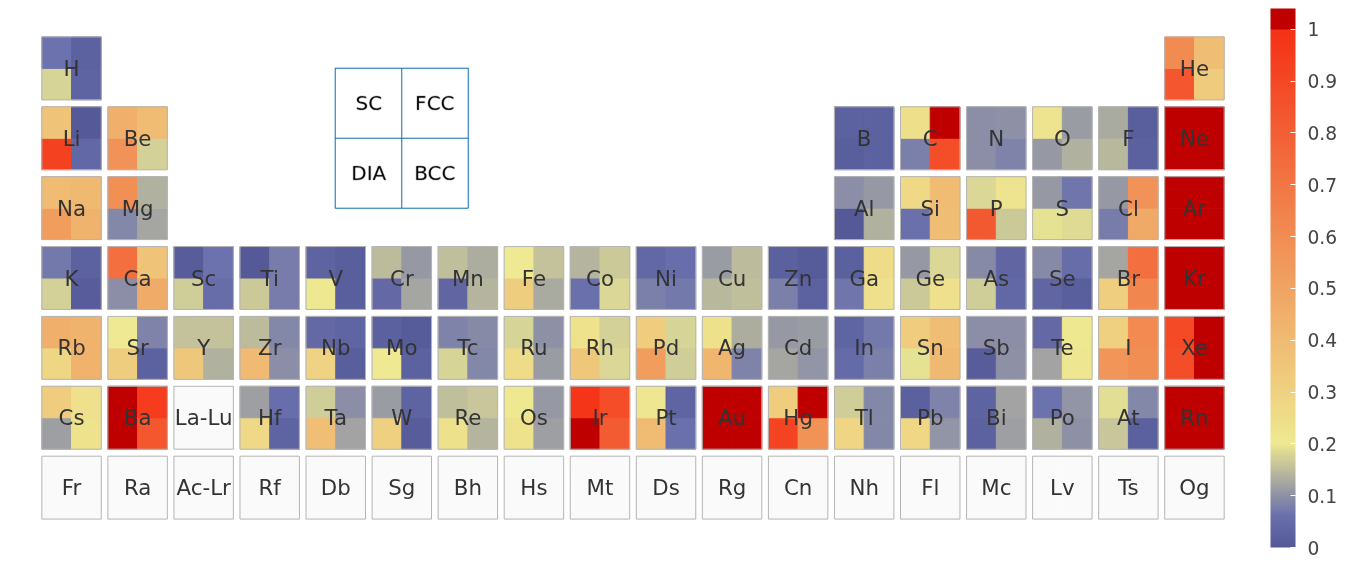}\\[0.5ex]
\includegraphics[width=\textwidth]{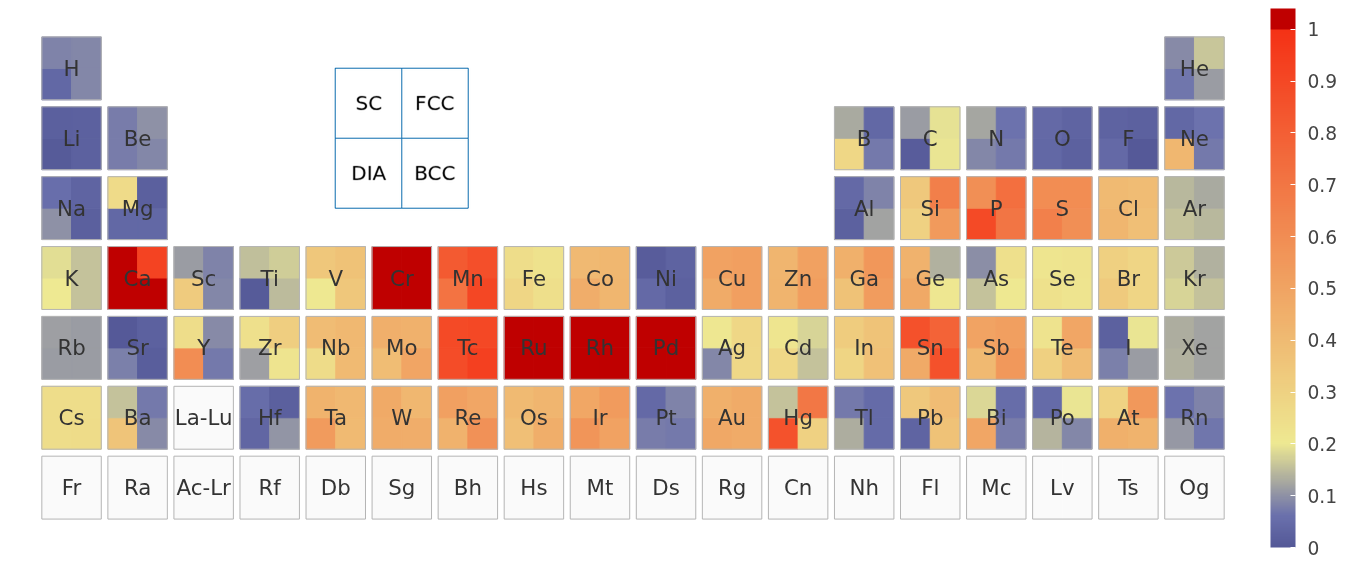}
\caption{\label{fig:pp_basis_old}
Error decomposition for the initial UZH settings. Top: CP2K-GTH-UZH relative to SIRIUS-GTH-UZH, which isolates the Gaussian-basis error. Bottom: SIRIUS-GTH-UZH relative to SIRIUS-FP-LAPW, which isolates the pseudopotential error.}
\end{figure*}

The decomposition in Eqs.~\ref{eq:decomposition_total}, \ref{eq:decomposition_basis}, and \ref{eq:decomposition_pp} separates two dominant sources of error, as shown in Fig.~\ref{fig:pp_basis_old}. First, the noble gases and elements such as Ba, Au, and Ir are basis limited. SIRIUS-GTH-UZH agrees well with SIRIUS-FP-LAPW, indicating that the GTH-UZH pseudopotential is transferable for these tests, but CP2K-GTH-UZH deviates from SIRIUS-GTH-UZH. These cases point to missing diffuse or polarization flexibility in the Gaussian basis for weakly bound or highly compressible solids. This observation is physically plausible because noble-gas equation-of-state curves are shallow and therefore sensitive to small basis-set and basis-set-superposition effects.

Second, for elements such as Ca, Cr, Pd, Rh, Ru, and Tc, SIRIUS-GTH-UZH differs substantially from SIRIUS-FP-LAPW, while CP2K-GTH-UZH remains close to SIRIUS-GTH-UZH. These cases are pseudopotential limited. Improving the Gaussian basis alone would not remove the dominant error because the same discrepancy is already present in the plane-wave representation of the pseudopotential.

\subsection{Improved UZH settings}

The classification in Table~\ref{tab:classification} was used to revise the settings in a targeted way. Basis-limited elements were treated by extending or reoptimizing the UZH-MOLOPT basis while retaining the condition-number penalty that prevents unstable diffuse functions. Pseudopotential-limited elements were treated by refitting the GTH parameters with the CP2K ATOM code, including DKH3 atomic reference information and semicore states where appropriate, followed by a MOLOPT basis reoptimization. This targeted route differs from a brute-force convergence exercise as it changes only the component that the error decomposition identifies as limiting. The central output of this step is therefore a revised set of CP2K parameter files, namely improved MOLOPT basis functions for basis-limited cases and improved GTH pseudopotentials and MOLOPT basis for pseudopotential-limited cases.

\begin{figure*}
\includegraphics[width=\textwidth]{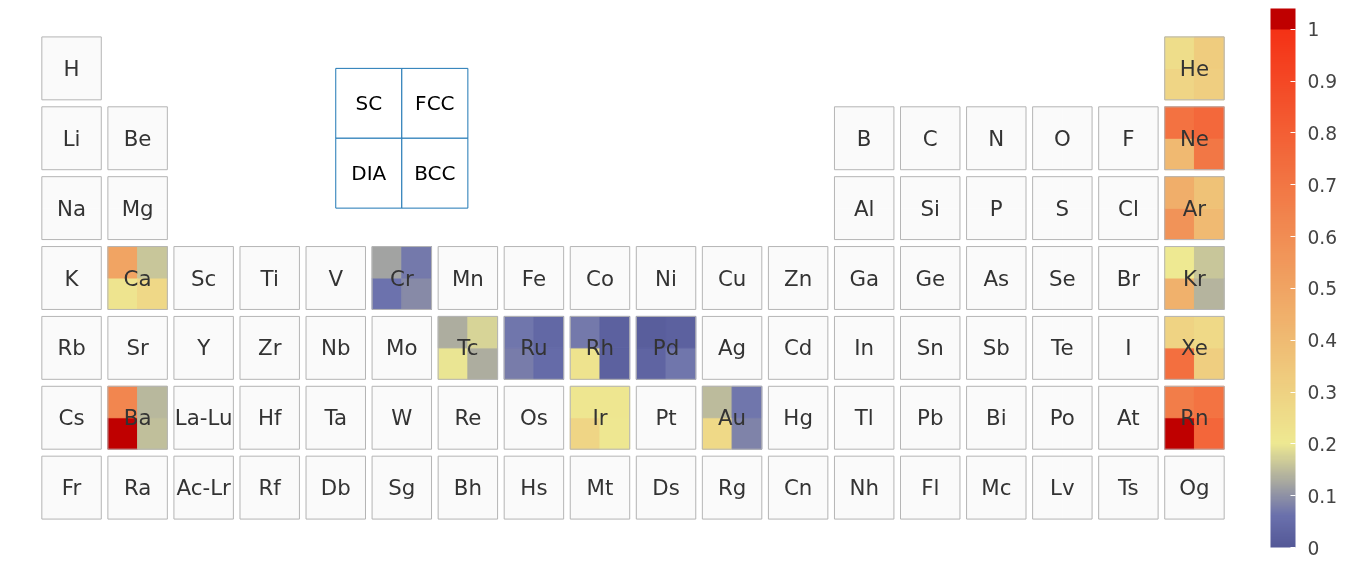}\\[0.5ex]
\includegraphics[width=\textwidth]{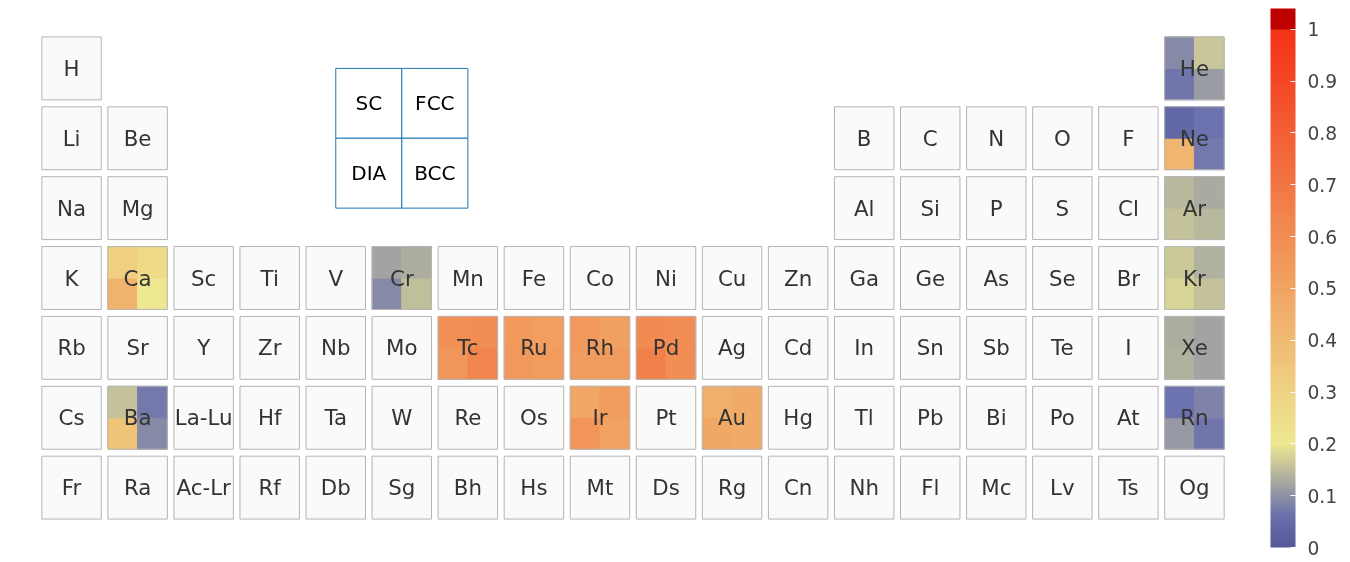}
\caption{\label{fig:component_updates}
Residual component-wise equation-of-state error after applying the UZH protocol. Top: CP2K-GTH-UZH-improved relative to SIRIUS-GTH-UZH-improved, which isolates the remaining Gaussian-basis contribution. Bottom: SIRIUS-GTH-UZH-improved relative to SIRIUS-FP-LAPW, which isolates the remaining pseudopotential contribution. The displayed elements form the recomputed improved subset; the actively revised component is element dependent, so basis-limited cases also appear in the lower panel as pseudopotential-transferability controls. White entries were not part of the corresponding residual comparison.}
\end{figure*}

Figure~\ref{fig:component_updates} reports the two residual component-wise comparisons for the improved settings. The upper panel compares CP2K-GTH-UZH-improved with SIRIUS-GTH-UZH-improved and therefore gives the remaining Gaussian-basis contribution. The lower panel compares SIRIUS-GTH-UZH-improved with SIRIUS-FP-LAPW and gives the remaining pseudopotential contribution. The displayed elements form the recomputed improved subset; this subset contains both basis-limited cases and pseudopotential-limited cases because both components must be evaluated on the same validation set. Thus, noble-gas entries in the lower panel are controls of pseudopotential transferability rather than indications that noble-gas pseudopotentials were refitted. At the level of the underlying signed equation-of-state energy differences, the two residual components combine to the remaining total CP2K-GTH-UZH-improved error relative to SIRIUS-FP-LAPW shown in the bottom panel of Fig.~\ref{fig:total_before_after}; the positive $\varepsilon$ values are component norms and should not be added entry by entry. Read together with the total-error reduction in Fig.~\ref{fig:total_before_after}, these maps show that the improvement is not a cosmetic redistribution of errors but a targeted reduction of the components identified as limiting. The supplementary material Figs.~6--10 show representative equation-of-state curves for Cr, Ba, Ne, Au, and Ga, connecting the periodic-table heat-map colors to the underlying energy-volume data and showing how basis-limited and pseudopotential-limited cases appear at the curve level.

The improved settings reduce the dominant outliers of the initial parameterization and make the remaining discrepancies more interpretable. In particular, the noble-gas errors are reduced primarily by basis-set changes, Ba is improved for the metallic prototypes but remains nontrivial in the diamond prototype, and the transition-metal cases respond mainly to pseudopotential refitting. The resulting UZH protocol is therefore neither a fixed table of parameters nor merely an error-reduction recipe. It is a constructive parameter-development workflow whose deliverable is an improved, validated set of CP2K basis-set and pseudopotential files.

\section{Implications for CP2K verification}

The UZH protocol turns CP2K verification into a diagnostic and constructive workflow that produces updated parameter files. The comparison with SIRIUS-FP-LAPW gives the total accuracy relevant for production calculations. The intermediate SIRIUS-GTH-UZH calculation tells whether the Gaussian basis or the pseudopotential is responsible. The CP2K machinery supplies the internal route for improvement with basis optimization with condition-number control, DKH3 atomic references, and GTH parameter fitting. This is particularly valuable for users who need reliable calculations across molecules and solids, because the protocol connects molecular transferability to periodic equation-of-state precision and closes the loop from verification to a distributable CP2K data set.

The protocol also clarifies what should be considered converged. Increasing the GPW grid cutoff alone cannot remove a Gaussian-basis error. Likewise, replacing a basis set cannot repair a pseudopotential whose scattering or semicore transferability is insufficient. Conversely, once the SIRIUS-GTH-UZH and SIRIUS-FP-LAPW curves agree, the remaining CP2K discrepancy is a basis issue that can be addressed without changing the pseudopotential. Tables~\ref{tab:epsilon-cp2k-uzh-merged-vs-sirius-fp-lapw} and \ref{tab:classification} make this point operational. The first reports the user-facing reduction of the total error, whereas the second identifies which numerical approximation should be targeted to obtain that reduction. This separation is the practical advantage of combining CP2K and SIRIUS in a single AiiDA-based verification framework.

\section{Conclusions}

We have presented the UZH protocol for constructing, diagnosing, and improving CP2K basis-set and pseudopotential settings. The protocol combines molecularly optimized MOLOPT basis sets with GTH separable dual-space Gaussian pseudopotentials, calibrates the basis on small molecules, and validates the resulting settings through AiiDA unary-crystal equation-of-state workflows. By comparing CP2K-GTH-UZH, SIRIUS-GTH-UZH, and SIRIUS-FP-LAPW, the total CP2K error can be decomposed into Gaussian-basis and pseudopotential components. The same CP2K infrastructure used for production calculations also provides the ATOM and basis-optimization tools required to improve the limiting component and to write the corresponding revised MOLOPT and GTH parameter files. The resulting workflow provides a reproducible path from verification data to systematically improvable CP2K simulations across molecules and condensed phases. The present UZH release provides MOLOPT Gaussian basis sets and GTH pseudopotential parameter files for three exchange-correlation levels: Perdew--Burke--Ernzerhof (PBE) at the generalized-gradient-approximation level, strongly constrained and appropriately normed (SCAN) at the meta-generalized-gradient level, and Perdew--Burke--Ernzerhof hybrid (PBE0) as a hybrid density-functional approximation with exact exchange~\cite{Perdew1996,Sun2016,Adamo1999}. We therefore propose the UZH protocol as a reproducible route for CP2K pseudopotential calculations and for corresponding all-electron verification calculations, including CP2K's GAPW approximation. The all-electron branch of the protocol and the accuracy and scope of the GAPW approximation will be discussed in detail in future work.

\section*{Supplementary Material}

See the supplementary material for CP2K ATOM, basis-optimization, and SIRIUS setup details; pseudopotential and MOLOPT basis-set parameter tables; molecular angle, dihedral, dipole, bond-class, and reference-code comparisons; representative equation-of-state curves; and the organization of the release repository.

\begin{acknowledgments}
The authors gratefully acknowledge the computing time made available to them on the high-performance computer OTUS at the Paderborn Center for Parallel Computing (PC2), a National High-Performance Computing (NHR) Center. PC2 is jointly supported by the Federal Ministry of Research, Technology and Space and the state governments participating in the NHR joint funding program (\url{www.nhr-verein.de/en/our-partners}). Part of the research was funded by the DFG (project numbers 398046241, 417590517/CRC1415 and 519869949).
\end{acknowledgments}

\section*{Author Declarations}

\subsection*{Conflict of Interest}

The authors have no conflicts to disclose.

\subsection*{Author Contributions}

Hossein Mirhosseini: investigation, data curation, visualization, writing--original draft, writing--review and editing. Tiziano M. A. M\"uller: investigation, data curation, software, validation. Matthias Krack: basis-set and pseudopotential curation, validation. Thomas D. K\"uhne: conceptualization, methodology, formal analysis, supervision, funding acquisition, writing--original draft, writing--review and editing. J\"urg Hutter: conceptualization, methodology, investigation, software, formal analysis, writing--review and editing.

\section*{Data Availability}

The data that support the findings of this study are openly available in the DCM-Uni-Paderborn GitHub repository \url{https://github.com/DCM-Uni-Paderborn/UZH-protocol}. The repository contains the official CP2K UZH basis-set and pseudopotential files, curated molecular-calibration structures, Gaussian~16 reference inputs and logs, CP2K molecular inputs and optimized-coordinate files, compact molecular summary tables, organized CP2K-GTH-UZH, SIRIUS-GTH-UZH, and SIRIUS-FP-LAPW equation-of-state result files, derived epsilon tables, per-element equation-of-state diagnostic plots, selected workflow input/output records, vendored ACWF verification and post-processing scripts, manuscript figures, source mappings, file inventories, repository citation metadata, licensing information, and SHA-256 checksums. The molecular-calibration subset was curated from the MolTest source record cited above, and the full upstream archive checksum is recorded in the repository provenance files. The supplementary material gives the corresponding directory layout.

\bibliography{bib_jcp_tdk}

\end{document}